\documentclass[apj]{emulateapj}

\bibpunct{(}{)}{,}{a}{}{;}

\shorttitle{Quantifying substructure}
\shortauthors{Starkenburg et al.}

\begin{document}

\title{Mapping the Galactic Halo VIII: Quantifying substructure}
\author{Else Starkenburg and Amina Helmi} 
\affil{Kapteyn Astronomical
Institute, University of Groningen, PO Box 800, 9700 AV Groningen, the
Netherlands \email{else@astro.rug.nl}}

\author{Heather L. Morrison and Paul Harding}
\affil{Department of Astronomy,
Case Western Reserve University, Cleveland OH 44106-7215}

\author{Hugo van Woerden}
 \affil{Kapteyn Astronomical Institute, University of Groningen, PO Box 800, 
9700 AV Groningen, the Netherlands} 

\author{Mario Mateo}
\affil{Department of Astronomy, University of Michigan, 821
Dennison Bldg., Ann Arbor, MI 48109--1090}

\author{Edward W. Olszewski}
\affil{Steward Observatory, University of Arizona, Tucson,
AZ 85721}

\author{Thirupathi Sivarani}
\affil{Department of Astronomy, University of Florida, Gainesville,
 FL 32611-2055}

\author{John E. Norris and Kenneth C. Freeman}
\affil{Research School of Astronomy and Astrophysics, ANU, Private Bag,
Weston Creek PO, 2611 Canberra, ACT, Australia}

\author{Stephen A. Shectman}
\affil{Carnegie Observatories, 813 Santa Barbara St, Pasadena, CA,
91101}

\author{R.C. Dohm-Palmer}
\affil{Department of Astronomy, University of Michigan, 821
Dennison Bldg., Ann Arbor, MI 48109--1090}

\and
\author{Lucy Frey and Dan Oravetz}
\affil{Department of Astronomy,
Case Western Reserve University, Cleveland OH 44106-7215}

\begin{abstract}

We have measured the amount of kinematic substructure in the Galactic
halo using the final data set from the Spaghetti project, a pencil-beam high-latitude sky
survey. Our sample contains 101 photometrically selected and
spectroscopically confirmed giants with accurate distance, radial
velocity, and metallicity information.  We have developed a new
clustering estimator: the ``4distance'' measure, which when applied to
our data set leads to the identification of one group and seven pairs of
clumped stars. The group, with six members, can confidently be matched to tidal
debris of the Sagittarius dwarf galaxy. Two pairs match the
properties of known Virgo structures. Using models of the disruption
of Sagittarius in Galactic potentials with different degrees of dark
halo flattening, we show that this favors a spherical or prolate halo
shape, as demonstrated by Newberg et al. using the Sloan Digital Sky Survey data. One
additional pair can be linked to older Sagittarius debris.  We find
that 20\% of the stars in the Spaghetti data set are in
substructures. From comparison with random data sets we derive a
very conservative lower limit of 10\% to the amount of substructure in
the halo. However, comparison to numerical simulations shows that our
results are also consistent with a halo entirely built up from disrupted
satellites, provided that the dominating features are relatively broad due
to early merging or relatively heavy progenitor satellites.
\end{abstract}


\section{Introduction and Outline}

In modern, cold dark matter dominated, cosmological models structure
builds up hierarchically, i.e., small structures collapse first and
then merge together to form larger structures. If such processes also
take place on galactic scales, we would expect to see merger debris in
the stellar halos of galaxies  \citep[e.g.,][]{bul05}. We should find substructures there that
remain coherent in phase space for many gigayears \citep{joh96,hel99}.
This prediction has led to intensive searches, especially in the Milky
Way, and to the development and exploitation of several surveys. These
include, for example, large photometric surveys like the Sloan Digital
Sky Survey \citep[SDSS][]{sdss07} and Two Micron All Sky Survey \citep[2MASS][]{2mass06}, but also
smaller surveys that use more distinct halo tracers like RR Lyrae
variables (e.g., QUEST, \citet{viv04}; SEKBO, \citet{moo03}), or
halo red giant stars such as the Spaghetti survey, first described in
\citet{mor00}.

These surveys have produced the much sought after observational
evidence for late merging in the outer halo of our Galaxy, which is
presumably associated with its hierarchical formation. The Magellanic
Stream \citep{mat74} and especially the Sagittarius dwarf galaxy
\citep{iba94,maj03} that is being tidally stripped by our Milky Way, are two
smoking gun examples. Other large-scale features found in the Galaxy
are the Monoceros stream, a relatively broad stream of stars of
debated origin \citep{new02,iba03,martin04,pen05} and several
substructures in the direction of Virgo
\citep{viv01,new02,zin04,jur08,duf06,new07,kel08}. Various small substructures 
have been found as well \citep[e.g.,][]{cle06,bel07a}, of which a remarkable 
example is the relatively narrow ``Orphan
Stream'' \citep{bel07b,gri06}. The existence of substructure is not
restricted to the Milky Way, but is also found in the stellar halos of
other galaxies \citep[e.g.,][]{sha98,deJ07,deJ08,mar08,mar09}, most 
notably in M31 where a prominent stellar stream
and a wealth of smaller tidal features have been detected
\citep{iba01a,iba07}.

Although many substructures have been uncovered \citep[e.g.,][]{bel06}, their role in the
formation of the Milky Way galaxy is still unclear. Is late
accretion a dominant or a minor factor in the buildup of the halo? Is
the halo dominated by a smooth component which underlies the
substructures we find?  Or do the discovered substructures represent
the tip of the iceberg and is the whole galaxy halo in fact the result
of merged (stellar) structures?

Most surveys carried out so far have been analyzed in a rather qualitative
manner, and so do not give a direct answer to these very fundamental
questions. The first thorough attempt at quantifying this process was carried
out by \citet{bell07}. They analyzed the amount of substructure in the
spatial distribution of the stellar halo using $\sim$4 million color-selected main-sequence
turnoff stars in the SDSS. The magnitude limits of this survey correspond to distances of 
$\sim$35 kpc from the Sun. They found that fractional rms deviations
on scales $\geq$ 100 pc from the best-fitting smooth halo model are
$\geq$ 40\%. Hence they concluded that the stellar halo is highly
structured, which is consistent with a scenario in which merging is an
important factor in the buildup of the halo.

In this paper, we statistically quantify the amount of substructure in
the halo, but now we combine spatial {\it and} kinematical data from
the Spaghetti survey. As we will show below, the addition of
kinematical data greatly improves our ability to identify
substructure. Additionally, our survey traces the halo using giant
stars to much larger distances of $\sim$100 kpc.  To achieve our goal,
we have developed a new substructure estimator.  This {\it 4distance}
estimator works in a four-dimensional space defined by the spatial
coordinates and radial velocities of stars. As we shall show below, it
particularly is suitable for finding substructures with similar sky
position, distance and radial velocities.

Our paper is organized as follows. In Section \ref{data set} we briefly
discuss the properties of the final Spaghetti data set (we defer a
more detailed description to H.L. Morrison et al., in preparation). In Section
\ref{4dist_sec} we present our substructure estimator and apply it to
this data set. Our results are compared with simulations of stellar halos
built up completely from accreted satellites in Section \ref{sec_sim}. In
Section \ref{newstructures_sec} we discuss whether any of the
substructures found in our analysis can be related to known
structures. We discuss and summarize our results in Section
\ref{sec_discuss} and \ref{sec_conc}.

\section{The data set}\label{data set}

The Spaghetti survey is a pencil-beam survey of high-latitude fields
that was completed in 2006. Washington photometry was used to
preselect red giant candidates \citep[see for more
details][]{mor00,doh00,mor01}. These candidates were then followed-up
spectroscopically as described in detail in \citet{mor03}.  By
measuring the strength of the \ion{Ca}{2} K, \ion{Ca}{1} $\lambda4227$
and Mg $b/$H features, metal-poor dwarfs and halo giants can be
distinguished in order to obtain a clean sample of K giants in the
halo.

All spectroscopically confirmed giant stars are included in the
Spaghetti data set. This final data set consists of 101 giants, from 13
separate spectroscopic runs. Two giants have distances $\geq$ 100 kpc,
33 of them have distances over 30 kpc. The typical errors on distance
are 15\%, on radial velocity 10--15 km s$^{-1}$ and on metallicity 0.25--0.3
dex. The sky coverage, distances, radial velocities, metallicities and
corresponding errors of the data set will be presented in a forthcoming
paper (Morrison et al., in preparation).

\section{The 4distance}\label{4dist_sec}

We expect debris from a merged satellite to remain spatially coherent
in the outer halo \citep[see, for example, the numerical simulations of
satellite accretion by][]{joh96}. Even when spatial structure is
no longer apparent, the debris from the merged satellite can still be
recognized in velocity space \citep{hel99}. Therefore, stars from the
same parent object should be clustered in phase space.

For the 101 giants in our data set we possess information on four of
the six phase space components: the spatial components (galactic
longitude, galactic latitude and distance), and radial velocity. With
these four components, we can define a measure of clustering by
computing a distance in a four-dimensional space for every pair of
stars in our data. We use \\ $l$ = galactic longitude,\\ $b$ =
galactic latitude,\\ $d$ = distance to the Sun,\\ $v$ = $v_{GSR}$ = line-of-sight velocity corrected for solar and local standard of rest (LSR)
motions,\footnote{We use a solar motion of ($v_{x}$, $v_{y}$, $v_{z}$) 
= (10.0, 5.2, 7.2) km s$^{-1}$ and $v_{LSR}$ = 220 km s$^{-1}$ \citep{deh98}}\\ $\phi$ = angular distance 
on the sky between the two stars. \\

We now define our 4distance between two stars $i$ and $j$ as follows: \\
\begin{equation}
4dist_{ij} = (w_{\phi}\phi^{2}_{ij} + w_{d}(d_{i} - d_{j})^{2} \nonumber \\
 + w_{v}(v_{i} - v_{j})^{2})^{0.5},  
\label{4disteq}
\end{equation}
where \\
$$\cos \phi_{ij} = \cos b_{i} \cos b_{j} \cos(l_{i}-l_{j}) +
\sin b_{i} \sin b_{j}.$$

Stars with small separations in this metric are expected to come from
the same object.

While the galactic longitude and latitude are incorporated as part of
the angular separation, the other components are used completely
independently in the final 4distance measure. The quantities
$w_{\phi}$, $w_{d}$, and $w_{v}$ are used to weigh the various
components, first normalizing by the range of this quantity (the
largest possible angular separation is $\pi$, distance 130 kpc and
velocity 500 km s$^{-1}$) and then by our observational errors on distance
($d_{err}$) and line-of-sight velocity ($v_{err}$). In the distance component, the
relative error $d_{err}/d$ is used\footnote{Here, the quantities within
$\langle \ \rangle$ denote the average errors over the whole sample.}
since distance errors scale with the distance itself:

\begin{equation}
w_{\phi} = \Bigl(\frac{1}{\pi}\Bigr)^{2}, \ \ w_{d} =
\Bigl(\frac{1}{130}\Bigr)^{2} \displaystyle \frac{\displaystyle
\Bigl(\frac{d_{err}(i)}{d(i)}\Bigr)^2 + \displaystyle
\Bigl(\frac{d_{err}(j)}{d(j)}\Bigr)^2}{2\Bigl\langle{\displaystyle
\frac{d_{err}}{d}}\Bigr\rangle^{2}}
\end{equation}
\begin{equation}
 w_{v} = \Bigl(\frac{1}{500}\Bigr)^{2} \, \displaystyle 
\frac{v^2_{err}(i) + v^2_{err}(j)}{2\langle{v_{err}}\rangle^{2}}.
\end{equation}

We find that the group-finding algorithm is quite insensitive to small
changes in the weighting factors. Multiple possibilities have been
explored, using several combinations of normalizing factors as well as
dependence of errors, which did not affect the key results presented
in this paper by more than a few percent.

\subsection{Choosing a relevant binsize}\label{sec_binsize}

We expect stars with small 4distance to be possible stream
members. However, the actual values of 4distance for stream members
will depend on a number of factors including the spatial sampling of
the Spaghetti survey.

\begin{figure}
\epsscale{1.00}
\includegraphics[width=\linewidth]{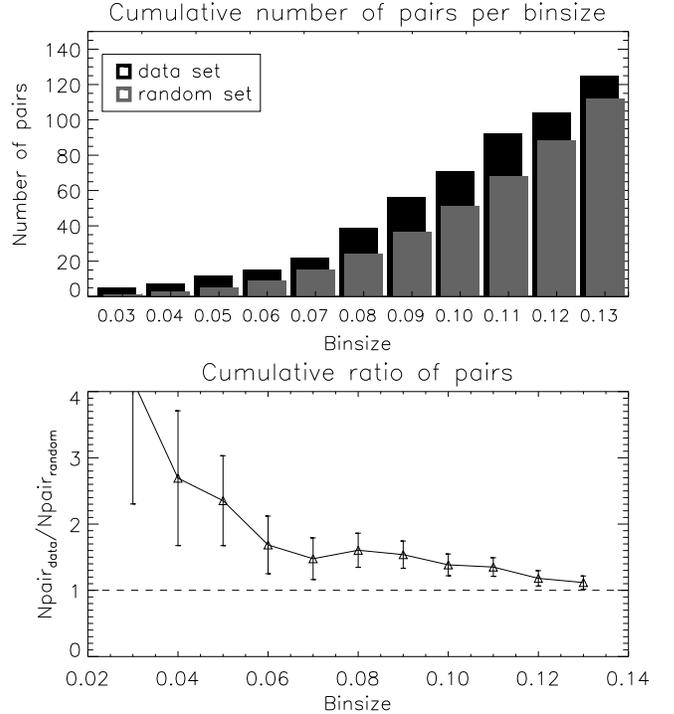}
\caption{Top panel shows the cumulative number of pairs found as a
function of $4dist$ for the Spaghetti data (black) and for the average
of 1000 random sets (gray). The bottom panel shows the cumulative
correlation function defined as the number of pairs in the data
divided by the average number of random pairs below a certain
$4dist$.\label{binsize}}
\end{figure}

We construct random samples in order to assess how often small values
of 4distance will occur by chance. To mimic our data set as much as
possible, we create random sets in which each star in the original
sample preserves its galactic longitude and latitude, but is randomly
supplied with a different (reshuffled) velocity and independently
with a different observed distance. In this way, we break any
correlations in the data while maintaining its global properties.  We
call two stars that are within a certain 4distance a
\textit{pair}. By comparing the total number of pairs at a certain
4distance in both the data and 1000 randomized data sets, the
significance in the number of pairs with small 4distance can be
investigated. This comparison is shown in Figure \ref{binsize}; the error
bars in the bottom panel are Poissonian.

The number of pairs within a certain 4distance measures the clumpiness
of the data at that particular scale. For all scales up to a $4dist$
of 0.13 plotted in Figure \ref{binsize}, the amount of clumpiness in
the data is larger than in the randomized set. Based on Figure
\ref{binsize}, we decide to investigate in more detail data pairs at
two different scales.  First, we would like to choose a $4dist$ within
which the ratio of data pairs to random pairs is sufficiently large
that our data pairs have a high chance of being real. Second, we
want to avoid throwing away pairs by being too restrictive.

We first focus on a $4dist$-scale of $0.05$. Table \ref{tab_4dist5}
lists what this scale corresponds to in physical units.\footnote{Assuming average error values.} For a pair of
stars with exactly the same radial velocity and distance from the Sun,
a $4dist = 0.05$ implies a maximum separation on the sky of
$9^{\circ}$. At a distance of 5 kpc this corresponds to a physical
size of 0.8 kpc, while at 50 kpc it would be 10 times larger. Note
however that the values given in the Table are clearly upper limits,
since no two stars will have the exact same values of the remaining coordinates.

For a $4dist \leq 0.05$, 12 pairs are found in the data set and on average
just $\sim 5$ pairs in the random sets. The bottom panel in Figure
\ref{binsize} shows that for $4dist \geq 0.05$ the significance
decreases. Supplementary to the core structures defined by the pairs found
at $4dist \leq 0.05$, we also explore if other stars in our data set have $4dist
\leq 0.08$ such that they could be added to these core structures.

\begin{deluxetable}{ll}
\tablecaption{Maximum separations in the metric for $4dist = 0.05$\label{tab_4dist5}}
\tablehead{\colhead{Maximum separation in} & \colhead{for $4dist=0.05$}}
\startdata
Angle on the sky & $9^{\circ}$ \\
Distance & 6.5 kpc\\ 
Radial velocity & 25 km s$^{-1}$ \\
\enddata
\end{deluxetable}

\subsection{Pairs and groups} 

Some of the 12 pairs with $4dist \leq 0.05$ can
be combined to form larger groups. We define
a group of stars when every member has a $4dist \leq 0.05$ with
\textit{at least} one other member of the group. This we call the
\textit{friends-of-friends} criterion. For each star$_{i}$ and
star$_{j}$ they belong to the same group if for:

\begin{equation}
\textrm{star}_{i} \ \exists \  \textrm{star}_{j} \ / \ 4dist_{ij} \leq \epsilon, \ \textrm{where} \ 
\epsilon = 0.05
\end{equation}

\begin{figure}
\epsscale{1.00}
\includegraphics[width=\linewidth]{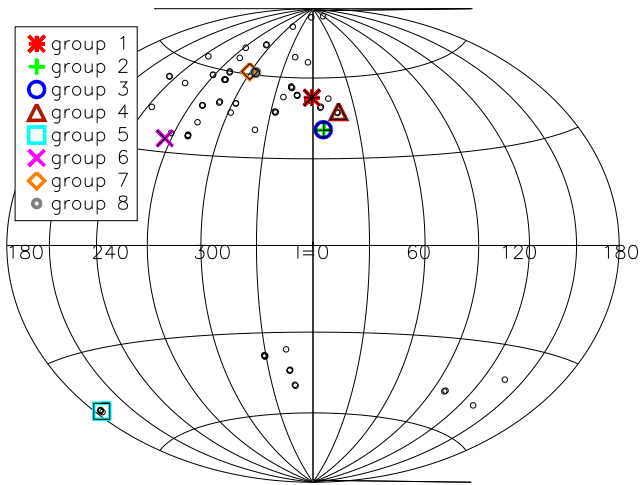}
\caption{Sky distribution of the data set highlighting the location of 
the group and pairs found. Note that pairs 2 and 3 are overlapping on the sky, 
but they possess very different velocities. \label{skyplot}}
\end{figure}

\begin{figure}
\epsscale{1.00}
\includegraphics[width=\linewidth]{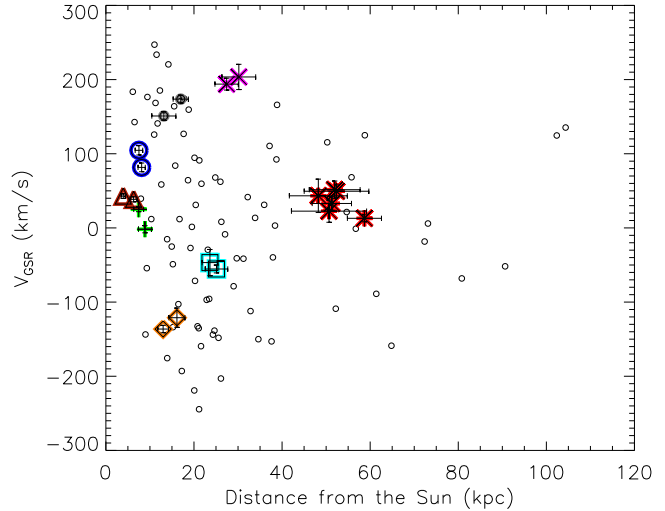}
\caption{Distribution of the data set in galactocentric velocity (defined 
as line-of-sight velocity corrected for the solar motion and LSR) 
vs. distance with the group and pairs overplotted. The color 
coding (in the online version) and the symbols are the same as in Figure \ref{skyplot}. \label{veldistplot}}
\end{figure}

\begin{figure}
\epsscale{1.00}
\includegraphics[width=\linewidth]{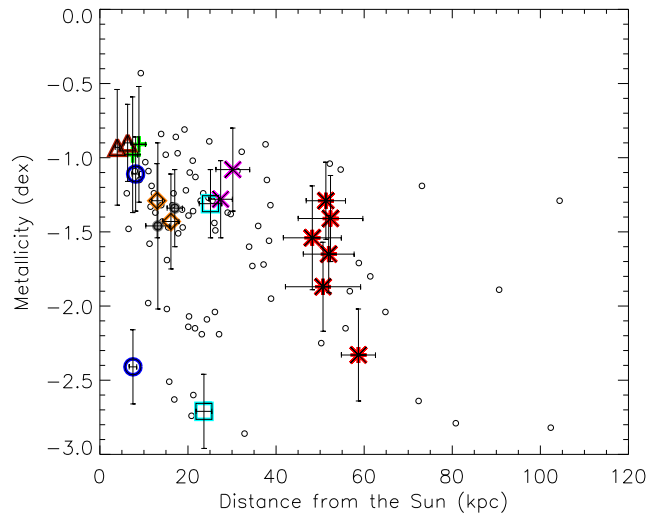}
\caption{Distribution of the data set in metallicity vs. distance 
with the group and pairs overplotted in different colors. The color 
coding (in the online version) and the symbols are the same as in Figure \ref{skyplot}.\label{metdistplot}}
\end{figure}

 Using this criterion one large group of five members is found. This
 leaves seven pairs that cannot be extended to groups with more
 members. We subsequently look at the added substructure within a
 $4dist$ of 0.08. All stars in the original core group of five are within
 a $4dist$ of 0.08 with every other member of the group, which
 strengthens the significance of this group. An extra member was found
 within a $4dist$ of $0.08$ of two of these stars. Therefore, this
 leaves us with one group of six stars and seven independent
 pairs. The properties of the final group and pairs are given in Table
 \ref{tab_final_groups} and are shown in Figures \ref{skyplot} (on the
 sky), \ref{veldistplot} (velocity vs. distance) and
 \ref{metdistplot} (metallicity vs. distance).

\begin{deluxetable*}{ crrrrcrrrr }
\tablewidth{0pt}
\tablecolumns{10}
\tabletypesize{\scriptsize}
\tablecaption{Positional, velocity and metallicity information for all giants in groups and pairs.\label{tab_final_groups}}
\tablehead{\colhead{} & \colhead{RA(2000)} & \colhead{DEC(2000)} & \colhead{$l$} & \colhead{$b$} & \colhead{[Fe/H]}  & \colhead{D$_{\odot}$} & \colhead{Vr$_{\odot}$} & \colhead{V$_{GSR}$} & \colhead{V$_{error}$} \\  & \colhead{h:m:s} & \colhead{$^{\circ}$:':''} &\colhead{(deg)} & \colhead{(deg)} & \colhead{(dex)} & \colhead{(kpc)} & \colhead{(km s$^{-1}$)} & \colhead{(km s$^{-1}$)} & \colhead{(km s$^{-1}$)} }
\startdata
1 &  15:15:27.06 &  +3:56:02.3 &    5.3037 & 48.5534 &       --1.41 $\pm$      0.29 &       52.36 $\pm$  7.34 &       23.7 &       49.5 &        8.6 \\
1 &  14:53:30.96  &	+1:25:42.2  &  356.8098 &  51.0609 &       --1.87 $\pm$    0.30 &       50.66 $\pm$       8.53 &       18.6 &       22.6 &       15.0  \\
1 &  14:52:42.98  &	+1:23:24.2 &    356.5443 & 51.1777 &       --1.29 $\pm$    0.26 &       51.30 $\pm$       4.46 &       29.6 &       33.0 &        9.2  \\
1 &  14:52:50.78  &	+1:29:46.6 &   356.7020 & 51.2284 &       --2.33 $\pm$    0.31 &       58.70 $\pm$       3.87 &        9.2 &       13.0 &        4.7  \\
1 &  14:52:40.14  &	 +1:03:18.8 &    356.1503 & 50.9520 &       --1.65 $\pm$    0.26 &       51.95 $\pm$       5.77 &       49.2 &       51.6 &       11.7  \\
1\footnote{This star is added to the group using the $4dist \leq 0.08$ criterion.} &  14:31:46.61 & +10:46:49.5 &    3.0593 & 61.2957 &  --1.54 $\pm$     0.35 &       48.23 $\pm$  6.58 &       26.4 &       43.3 &       22.5  \\
\tableline
2 & 15:44:49.80 &  --0:22:12.8 &     6.7604 & 40.1085 &      --0.91 $\pm$       0.39 &       8.87 $\pm$ 1.47 &      --34.3 &      --1.8 &         4.8  \\
2 & 15:44:53.56  & +0:05:52.4 &     7.2636 & 40.3766 &      --0.98 $\pm$       0.39 &        7.44 $\pm$  1.12 &       --8.8 &       25.1 &        3.2  \\
\tableline
3 &  15:44:45.29 & --0:17:03.4 & 6.8354 & 40.1752 &       --1.11 $\pm$       0.25 &      8.12 $\pm$  0.82 &       48.9 &       81.6 &        5.9  \\
3 & 15:42:47.40 &  --0:07:00.2 &     6.6228 & 40.6683 &       --2.41 $\pm$       0.25 &      7.51 $\pm$  0.86 &       72.7 &       104.6 &       6.3  \\
\tableline
4 & 15:39:05.02  & +10:28:36.5 &     18.2709 & 47.2382 &      --0.90 $\pm$      0.26 &        6.30 $\pm$    0.78 &      --21.1 &       38.6 &        3.4  \\
4 & 15:40:23.26 &  +10:13:46.6 &     18.1812 & 46.8380 &      --0.93 $\pm$      0.39 &        3.94 $\pm$    0.46 &      --16.9 &       42.9 &        3.2  \\
\tableline
5 &     3:26:11.85 &	--2:29:40.1 & 186.0313 & -45.5498 &       --2.71 $\pm$    0.25 &       23.62 $\pm$       1.78 &      --18.1 &      --46.8 &       17.7  \\
5 & 3:25:31.69 &	--1:45:08.8 &     185.0501 & -45.2278 &       --1.31 $\pm$    0.23 &       25.11 $\pm$       2.56 &      --29.3 &    --55.4    &      5.3   \\
\tableline
6 & 10:34:00.09 &   --19:00:13.8 &     263.3616 & 33.1049 &       --1.28 $\pm$    0.26 &       27.42 $\pm$       2.66 &      378.3 &       193.9 &       8.0  \\
6 &  10:54:02.91 &   --19:01:19.0 &    268.0987 & 35.7882 &       --1.08 $\pm$    0.28 &       30.18 $\pm$       3.83 &      382.2 &       203.6 &      17.0  \\
\tableline
7 &  12:32:11.52 & 	--1:03:44.5 &    292.8318 & 61.4314 &       --1.29 $\pm$    0.25 &       13.00 $\pm$       1.22 &      --45.3 &      --136.4 &       5.0  \\
7 &  12:56:08.49 &   	--2:16:23.8 &    305.3242 & 60.5766 &       --1.43 $\pm$    0.32 &       16.13 $\pm$       1.82 &      --39.9 &      --121.1 &      13.0  \\
\tableline
8 & 12:54:54.56  & 	--2:20:29.4 &     304.6941 & 60.5184 &       --1.34 $\pm$    0.26 &       17.00 $\pm$       1.73 &      255.7 &       173.6 &       5.4  \\
8 & 12:56:14.74  & 	--1:30:17.8 &     305.4380 & 61.3434 &       --1.46 $\pm$    0.56 &       13.16 $\pm$       2.73 &      229.8 &       150.9 &       6.0  \\
\enddata
\end{deluxetable*}

\subsection{Significance of the group and pairs}\label{siggroup_sec}

We now investigate the significance of the group and pairs. At both
levels, $4dist \leq 0.05$ and $4dist \leq 0.08$, more substructure is
found in the data set compared with the random sets. The core group of
five members, found at the $4dist \leq 0.05$ level, stands out very
significantly. We find a probability of about 1\% to obtain such a
large group in our random sets. The probability of finding pairs in
our random sets is significantly higher, in almost all random sets at
least one pair is found, but only in $\sim$1\% of our random sets we
find the same high fraction of stars to be paired. This implies that,
while finding some pairs in a random set has a high probability,
finding 19 stars in pairs at a level of $4dist \leq 0.05$, as we do in
our data set, is a highly improbable event.

The metallicity of the stars is not used as a criterion to select pairs. While a certain spread in metallicity can be expected in
stars originating from a disrupted satellite, the observed spread
gives additional information about the probability that a group or pair is
real. In our case especially pair 3 and pair 5 show
a large range in metallicity, as can be seen in Figure \ref{metdistplot}. 
For the other pairs however, the fact that their members are close in 
metallicity as well as in sky position and radial velocity makes them, 
and our selecting algorithm, more credible.

\section{Simulated data sets}\label{sec_sim}

We would like to constrain what fraction of the stellar halo has been
built from accreted satellites using the results from the previous
section. To this end, and to test the 4distance method used, we
compare our data set to a simulation of a halo which is entirely the
result of disrupted dwarf galaxies. For this purpose we use the
simulations from Harding et al. (2001)\nocite{har01} that model the
destruction by the Milky Way galaxy of a $10^{7} M_{\odot}$ satellite 
on different orbits.

\begin{figure}
\epsscale{1.00}
\includegraphics[width=\linewidth]{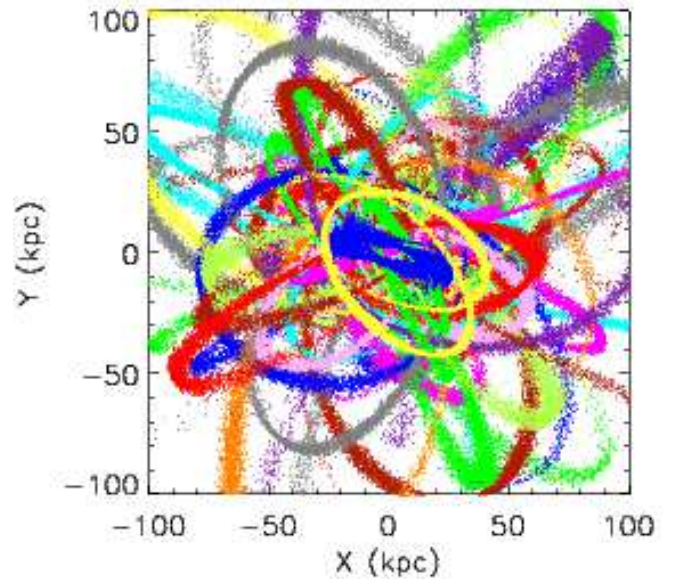}
\caption{X -- Y projection showing the distribution of streams in our
simulated stellar halo built up from 40 disrupted
satellites. \label{sim_xy}}
\end{figure}

Forty of the original satellite simulations were re-sampled so that
each particle corresponds to one halo K giant. This leaves nearly 8700
particles per satellite. The distribution of these K-giant simulations
after 10 Gyr is shown in Figure \ref{sim_xy}.

To create a halo built up completely out of disrupted galaxies, the
endpoints of the simulations (i.e., evolved for 10 Gyr) are used. From
this sample of over 340,000 simulated `giants' we draw subsets of 101
stars by requiring that the observed sky distribution, distance and
radial velocity distribution of the Spaghetti data set are matched within 
certain binsizes. In this manner, 30 simulated data sets are drawn which 
closely resemble the Spaghetti observations.

\subsection{Substructure in the simulated data sets}\label{sec_4dist_sim}

We look for substructure in the simulated data sets using the 4distance
method. In order to make a fair comparison, the simulated `giants' are
convolved with errors, to mimic the observational uncertainties. For
the distance a relative error of 15\% is used, while the velocities
are convolved with the same errors as found in the Spaghetti data set.

\begin{figure}
\epsscale{1.00}
\includegraphics[width=\linewidth]{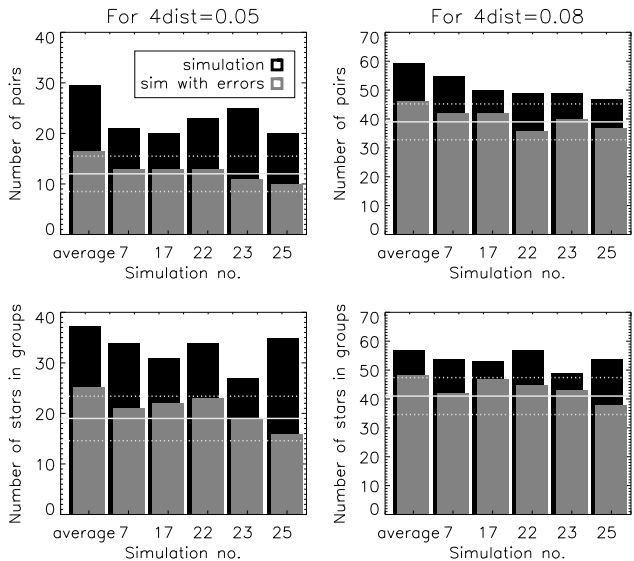}
\caption{Top panels show the number of pairs detected with the 4distance method at $4dist \leq 0.05$
and $4dist \leq 0.08$ for the average of 30 ``K-giant'' simulations constructed with the
sampling of the Spaghetti survey and for the five simulated sets out of these simulations that
most resemble the Spaghetti data set in terms of the amount of
substructure found. The black and gray bars correspond to the results
obtained for the simulated subsets before and after (observational)
error convolution. The white horizontal line represents the number of
pairs found in the Spaghetti data set, the dashed lines show 1$\sigma$
deviation assuming Poissonian statistics. The bottom panels show, for
the same simulations, the number of stars associated with substructures
according to our algorithm.\label{sim_groups}}
\end{figure}

 In Figure \ref{sim_groups}, we compare the number of pairs found below
$4dist$ of 0.05 and 0.08 within the simulated sets to the Spaghetti
data set. On small scales ($4dist = 0.05$), the average number of pairs
in the 30 simulated sets is significantly higher than that in the Spaghetti
data set. This effect is slightly less on a $4dist=0.08$ scale. This
implies that the simulations contain too much small-scale structure
compared with the data. Just five of the
simulated data sets show a similar amount of substructure as the
observed sample.

This implies that the simulated data sets have a significant fraction
of stars originating in narrower streams than in the observed data set. 
Because these streams are more confined, this substructure is easily 
picked up by our 4distance method. The fact that nearly all simulated 
data sets show more substructure especially at $4dist \leq 0.05$ means 
that the structures we trace in the Milky Way halo are broader. 

In the simulations, the average number of stars associated with pairs
according to our algorithm is 25\%, despite the fact that all of the
halo was built by accretion. This points to a fundamental limit in our
ability to recover substructure which is due to the rather small
sample of stars we have at our disposal. If the number of `giants' in
each of our simulated data sets were to be increased to $\sim1.000$,
the $4dist$ method would find on average 76\% of the `stars' to be in
substructures. Clearly, larger spectroscopic surveys should thus be
able to improve significantly on the limits we have set with the
Spaghetti project.

\subsection{Performance of the 4Distance method}\label{sec_4dist_sim2}

In this section we describe several tests performed using the
simulations to test the reliability and determine the strengths and
weaknesses of the 4distance method. First, we now check in the
simulated data sets whether the substructure found originates from a
common parent satellite, i.e., whether all pair members originate in
the same progenitor. At $4dist \leq 0.05$ 76\% of the pair members in
all the simulated data sets do share common parent satellites (this
number grows to 87\% if we do not take observational errors into
account). The number of mismatches is slightly larger when we look at
$4dist \leq 0.08$, but still 64\% of the pair members at this level
share common progenitors. As we increase the numbers of the `giants'
per bin in the simulated data sets by a factor of 10 to $\sim1.000$ in
total, 81\% of the pair members (convolved with errors) at $4dist \leq
0.05$ do share a common parent satellite. These percentages show that
the 4distance works well in the sense that it does not produce many
``false positives'': pairs that do not originate from a common
progenitor.

\begin{figure}
\epsscale{1.00}
\includegraphics[width=\linewidth]{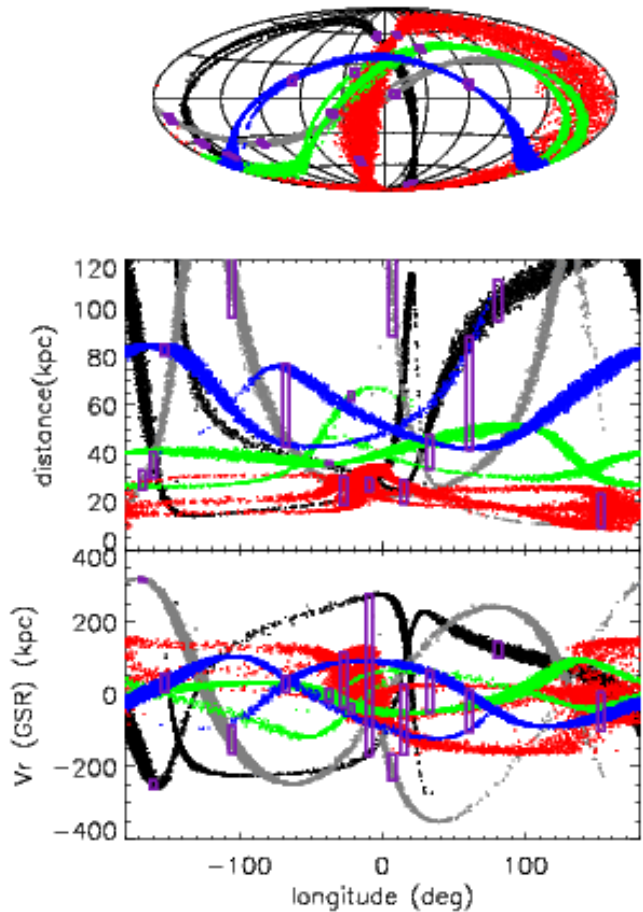}
\caption{All `giants' from five simulations of disrupted
satellites (in the online version the five satellite streams have
different colors) with on top 15 boxes of 5$^{\circ}$ by 5$^{\circ}$
(purple in the online version) we selected on the streams. Top panel:
the `giants' plotted on the sky in galactic longitude (centred at
zero) and galactic latitude. Middle and bottom panel: the same
simulations and boxes in galactic longitude vs. distance from the
Sun and radial velocity, respectively.\label{sim_skydistrstreams}}
\end{figure}

\begin{figure*}
\includegraphics[width=\linewidth]{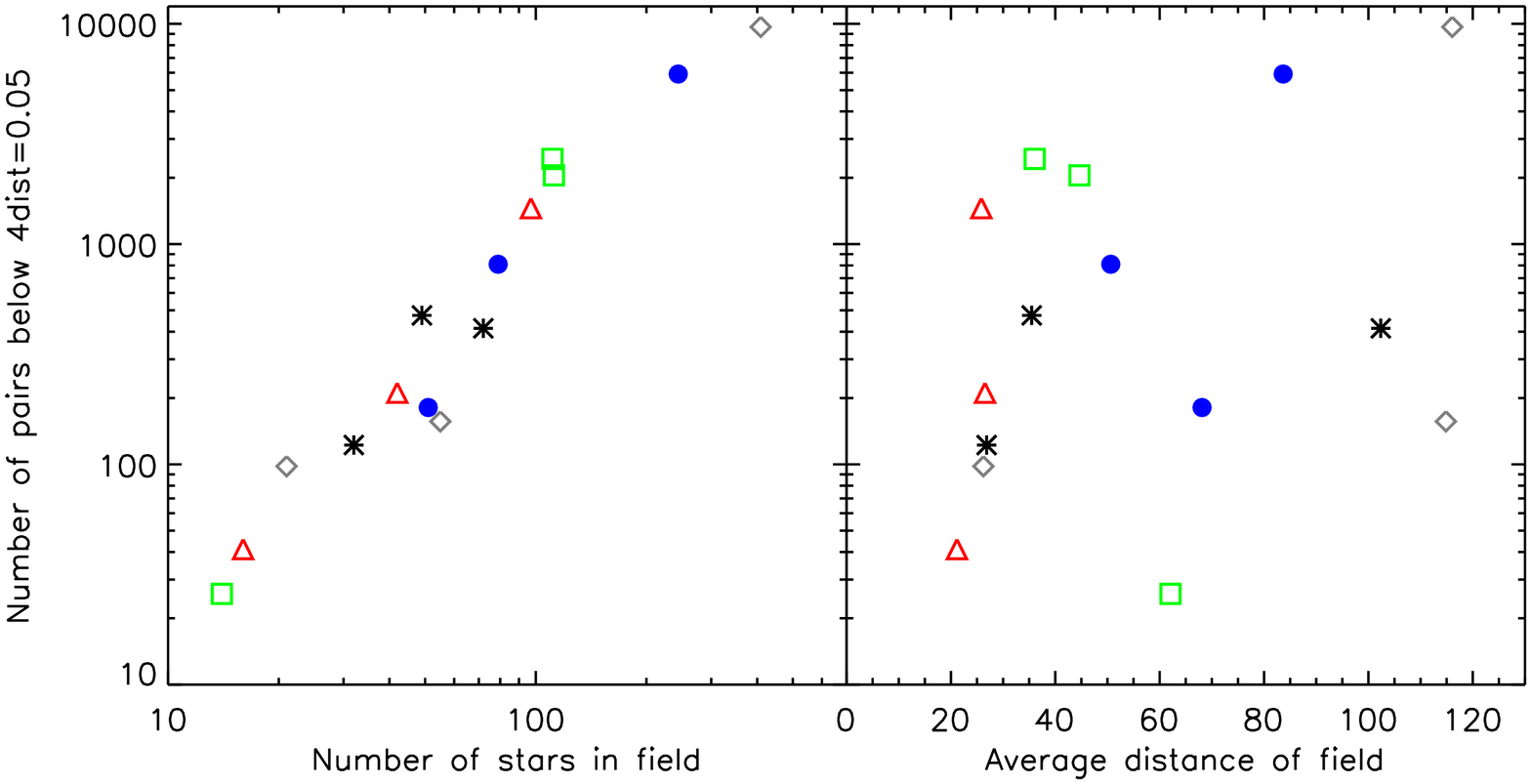}
\caption{Left panel: the number of pairs found in each of the 15
fields shown in Figure \ref{sim_skydistrstreams} as a function of the
number of `giants' selected in that field. Fields from the same
satellite stream have the same symbol (and the same color as in Figure
\ref{sim_skydistrstreams} in the online version). The number of giants
in the field is directly proportional to the surface brightness of the
stream.  Right panel: the number of pairs found in each selected field
as a function of the average distance of these giants from the
Sun. \label{4dist_simsurfaceB}}
\end{figure*}

To further explore the effectiveness of the 4distance method we have
selected five disrupted satellites from our simulations. Their debris
streams have different surface brightness and they move on different
orbits, with distances to the Sun that range from less than 10 kpc to
more than 120 kpc. We use this subset of the simulations to test which
of these characteristics determine how well the 4distance method
performs as measured by the number of pairs with $4dist \leq
0.05$. For this purpose, we select 15 fields of 5$^{\circ}$ by 5$^{\circ}$, three on each stream. The properties of the fields are shown
in Figure \ref{sim_skydistrstreams}. The number of stars in each field
is clearly different, corresponding to a difference in surface
brightness. In Figure \ref{4dist_simsurfaceB} we plot the number of
pairs at $ 4dist \leq 0.05$ found in each field, as a function of the
stars in each field (left panel) and as a function of the average
distance of the stars in the selected field (right panel). This figure
shows that the average distance has little (if any) bearing on the
number of pairs found by the 4distance algorithm. Clearly, the most
important factor determining the number of pairs found is the number
of giants (i.e., surface brightness) of the stream.

The surface brightness of a stream depends on several factors
\citep[see, e.g.,][]{hel99,joh01}:
\begin{itemize}
\item{the orbit (more extended orbits give rise to streams with a lower
surface density)} 
\item{the initial mass and phase space density of the progenitor
system (for a fixed mass, denser systems give rise to streams with
higher surface brightness, while at fixed density less massive objects
give rise to streams with lower surface density)}
\item{the time since formation of the stream, compared to the orbital
period (older accretion events produce lower surface brightness streams by the
present day) \citep[see also][]{joh08}}. 
\end{itemize}
Therefore the success of the 4distance method depends on all these
factors (indirectly) because they all impact the surface brightness of
a stream at the present day, but it is only the surface brightness of
the stream which affects its ability to recover substructure.

We use the same subset of five simulations to demonstrate how crucial
(radial) velocity information is to identify streams, since this is a
key feature of our survey and of ongoing projects such as the SEGUE
K-giant survey \citep{yan09}. Although the power of additional
velocity information can already be seen in Figure
\ref{sim_skydistrstreams}, we quantify this advantage here. For this
purpose we select a 5$^{\circ}$ by 5$^{\circ}$ field which stretches from 2$^{\circ}$ to 7$^{\circ}$
in galactic longitude and 35$^{\circ}$--40$^{\circ}$ in galactic
latitude in Figure \ref{sim_skydistrstreams}, and which contains 134
`giants'. In this particular field, four different streams cross
each other, providing us with an opportunity to demonstrate the value
of additional radial velocity information in such cases, which are
known to occur in the Milky Way halo as well (e.g., the streams from
Sagittarius near the North Galactic Pole).

\begin{figure}
\epsscale{1.00}
\includegraphics[width=\linewidth]{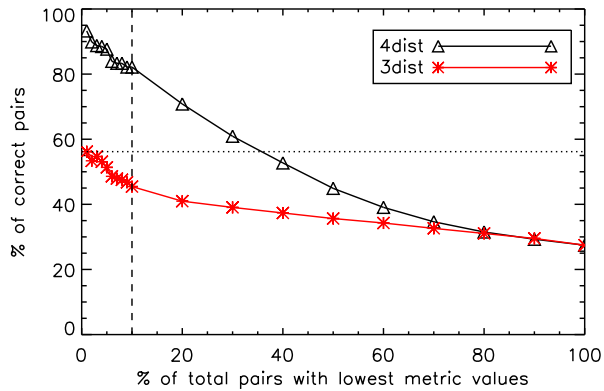}
\caption{For the 5$^{\circ}$ by 5$^{\circ}$ field taken from Figure
\ref{sim_skydistrstreams} around $(l,b) = (5^{\circ}, 38^{\circ})$
this figure shows the percentage of `correct' pairs found with the
4distance or 3distance method as function of the percentage of the
total number of pairs found, sorted by their metric (4distance or
3distance) value. The vertical dashed line therefore indicates the
10\% of pairs in the sample which have the lowest metric (3distance or
4distance) value (in this example these are all the pairs with a
$4dist$ value of $\leq 0.05$, of which more than 80\% are correct).
The horizontal dotted line indicates that the percentage of correct
pairs when the smallest 1\% 3distance values are considered is already less than
60\%. \label{4dist_simoverlap}}
\end{figure}

Subsequently we calculate both the 4distance value for all pairs from
the 134 `giants' in this field and also a `3distance' value, omitting
the velocity term in Equation \ref{4disteq}. Although the absolute
values will vary for both methods, we can still make a fair comparison
by comparing the most significant pairs in both methods. These are
defined as those with the smallest metric values. In Figure
\ref{4dist_simoverlap} we plot the percentage of `correct' pairs
(pairs for which the members originate from a common parent satellite)
for a fixed number (expressed as a percentage of the total number) of
most significant pairs sorted in increasing order of the metric values
for each method. From this figure it is very clear that the 4distance
method performs much better in selecting correct pairs at small
4distances, whereas the 3distance method is picking up many false
positives already for its most significant pairs. For comparison, the
$4dist \leq 0.05$ restriction used on the Spaghetti data set in Section
\ref{sec_binsize}, corresponds to $\sim10\%$ of the total number of
pairs (shown as the vertical dashed line). With the 4distance method,
the percentage of correct pairs at this level is over 80\%, while
without the velocity information one would select roughly 50\% false
positive pairs. This shows that velocity information is crucial in
identifying individual streams, especially because several streams can
be overlapping on the sky.

We also tested the possibility to link the found substructures with
the 4distance method to larger, more streamlike features using a
great circle counts method \citep{lyn95,pal02}. We found however that
such an approach is not suitable for our data set because of the pencil-beam character of our survey. The method and discussion of the results
can be found in the Appendix. In the next section we explore the
performance of the 4distance method in recovering known substructure.

\section{Are these substructures new?}\label{newstructures_sec}

Using the 4distance method, we identified one group and seven pairs of stars which are
likely to be real substructures in the Milky Way halo. We now explore
whether these can be related to other structures previously
discovered. 

\subsection{The Sagittarius Dwarf Galaxy}\label{sec_sag}

In \citet{doh01}, the Spaghetti Collaboration reported a concentration
of giant stars which stood well above the expectations of a smooth
halo model. In particular, four stars could be matched to a simulation
model of the debris from the disrupting Sagittarius spheroidal dwarf
galaxy \citep{hel01}.

In our examination of the full Spaghetti data set we find this same
overdensity to be very prominent: our largest group, with six
members. The substructure, at $l = -3.8^{\circ}$ to $ l =
5.3^{\circ}$, $b = 48.6 - 61.3^{\circ}$ and distances between 48 and
59 kpc, has properties in excellent agreement with the debris
predictions of models from Helmi (2004)\nocite{hel04}. Several other 
studies have reported overdensities in this region of the sky and at 
similar distances \citep[e.g.,][]{yan00,ive00,iba01b,mar01,viv01,maj03,sir04,bel06} 
which have also been associated with debris from Sagittarius. At this region
in space several wraps of Sagittarius cross each other, some recently
stripped from the satellite and some stripped quite
early.\footnote{These multiple wraps are observed in all models,
whatever the assumed shape of the dark halo.} This raises the
possibility that the stars do not all originate from the same wrap which 
can explain the spread in metallicities observed. Our method did not
find Dohm-Palmer et al. (2001)\nocite{doh01}'s additional proposed
structures at distances of 20 and 80 kpc to be significant, 
because the stars in these structures do not define a tight clump on the sky.

The average metallicity of the giants in group 1, [Fe/H]$\approx
-1.7$, is lower than the mean and about $-1.0$ dex lower than is found
for stars in the leading arm closer to the main body of the
Sagittarius dwarf galaxy \citep[e.g.,][]{mon07, cho07}. Because the
stars in the outskirts of a galaxy are stripped off first, a
metallicity difference between wraps would then reflect a metallicity
gradient in the dwarf galaxy itself. Dwarf spheroidal galaxies do in
fact often possess metallicity gradients
\citep[e.g.,][]{tol04}. Furthermore, a strong difference in horizontal
branch morphology has been detected between the Sagittarius core and a
portion of the leading arm of the Sgr stream \citep{bela06}. This
difference is consistent with the difference in metallicity between
the core and the stream stars in our group 1: the metal-richer core
has more red horizontal branch stars and the metal-poorer stream has
more blue horizontal branch stars.

Also pairs 5, 7, and 8 could be associated with the disrupting dwarf
galaxy. Comparison to simulations of Sagittarius by \citet{hel04}
shows this is possible if these stars were stripped off early, between
3 and 6 Gyr ago. Pair 5 matches best in a prolate halo. The membership
of pairs 7 and 8 is, however, more likely if the galactic halo
potential is significantly more oblate than prolate (when the
flattening of the potential is $q= 0.8$), but we think there is
stronger evidence these pairs are linked with the Virgo overdensities
and not to Sagittarius debris (see the next section). The group and three
pairs and a Sagittarius prolate model are plotted on the sky, in
distance and velocity in Figures \ref{SV_polar1}--\ref{SV_distvel}.

\subsection{The Virgo Substructures}\label{sec_virgo}

Several overdensities have been discovered toward the constellation
of Virgo \citep{viv01,new02,zin04,jur08,duf06,new07,kel08}. The origin 
of these features and whether they are all part of the same large structure 
is still unclear \citep{new07}. Two substructures (pairs 7 and 8) are near 
these overdensities on the sky and have distances that agree with those
measured for the Virgo overdensities. Also the velocities, although
very different for the two pairs, match approximately those reported
in the literature.

\begin{figure}
\epsscale{1.00}
\includegraphics[width=\linewidth]{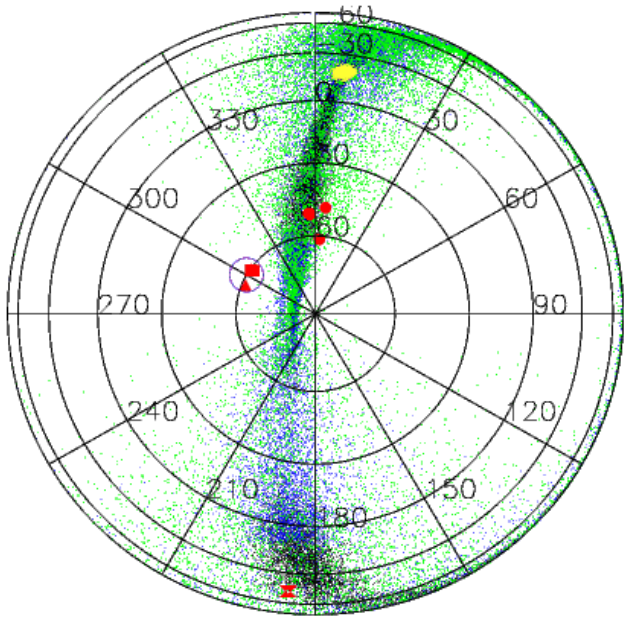}
\caption{Polar plot in galactic coordinates (l,b) of the whole sky
showing the location of debris from a model of the Sagittarius dwarf
galaxy orbiting in a prolate dark halo (q=1.25) potential. The debris
is color-coded (in the online version) according to ``dynamical age'': stripped off less than
3 Gyr ago (black dots), between 3 and 6 Gyr ago (blue dots), more than 6 Gyr
ago (green dots). The main body of the galaxy is shown in yellow. Also
overplotted are group 1 and pairs 5, 7, and 8 (red circles, hourglasses,
triangles, and squares, respectively) and the approximate region of the
overdensity found in SDSS photometry by \citet{new07} (purple
circle).\label{SV_polar1}}
\end{figure}

The same two pairs were also discussed in the previous section as
possible matches with older Sagittarius debris. This is not surprising
since this debris is close to Virgo on the sky. \citet{new07} have
however convincingly shown on the basis of SDSS photometry that the
Virgo overdensities are too low in latitude to be members of the
(leading) tidal tail of Sagittarius. Indeed in prolate halo models it
is predicted that the Sagittarius stream should not overlap with the
Virgo overdensities, as shown in Figure \ref{SV_polar1}.

\begin{figure}
\epsscale{1.00}
\includegraphics[width=\linewidth]{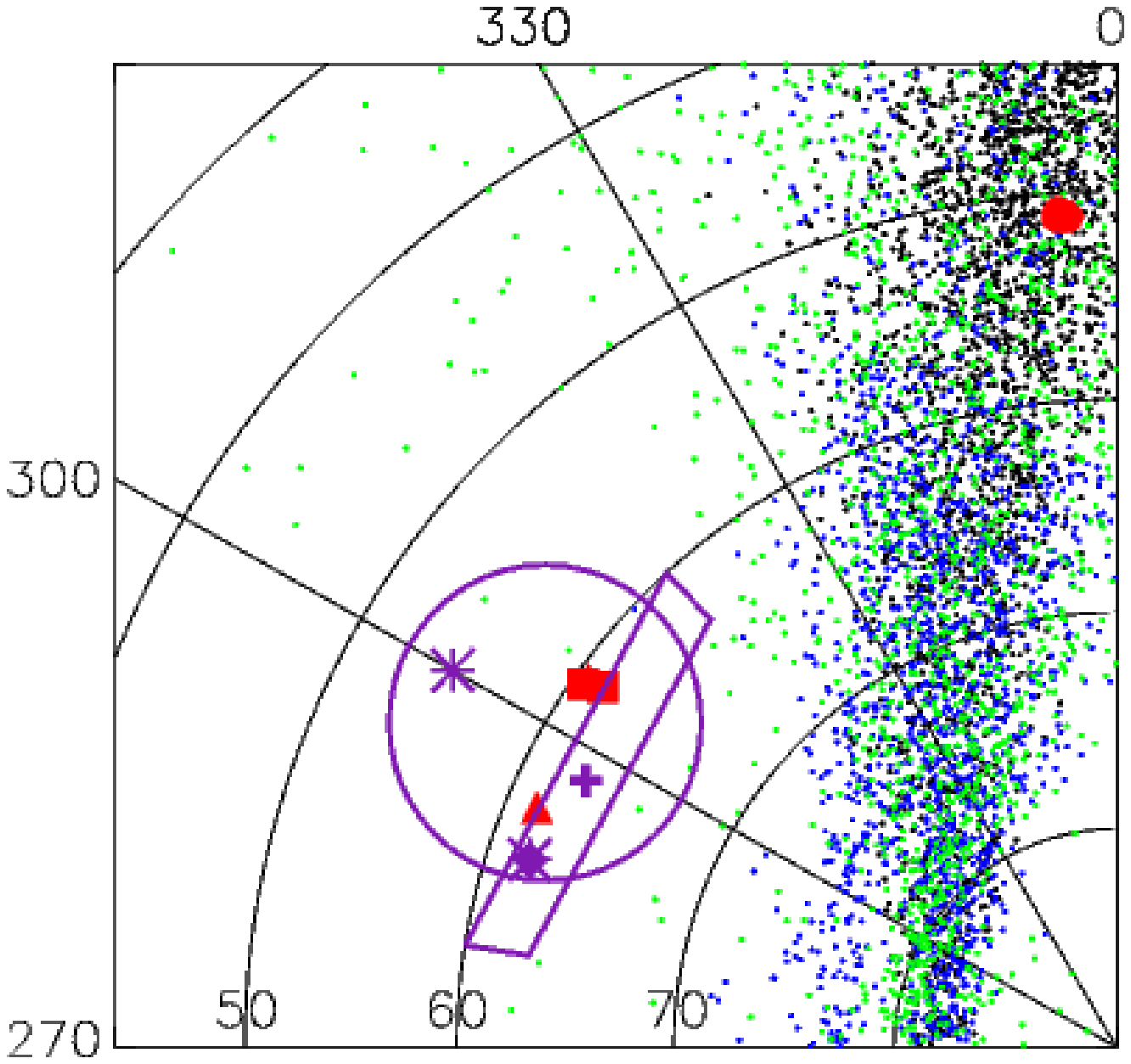}
\caption{Zoom-in of Figure \ref{SV_polar1} onto the Virgo
region. The same color coding (in the online version) and symbols are used as in Figure
\ref{SV_polar1}. The additional rectangular shape denotes the region of the RR
Lyrae overdensity in QUEST \citep{zin04}. A purple diamond shows the
position of the Virgo Stellar Stream \citep{duf06}, a purple plus
the overdensity S297+63-20.5 in SDSS \citep{new02} and the purple
asterisks point at two directions in which spectra are obtained with
SEGUE \citep{new07}. One of the SEGUE plates overlaps with the
direction of the Virgo Stellar Stream on the sky. \label{SV_polar2}}
\end{figure}

\begin{figure}
\includegraphics[width=\linewidth]{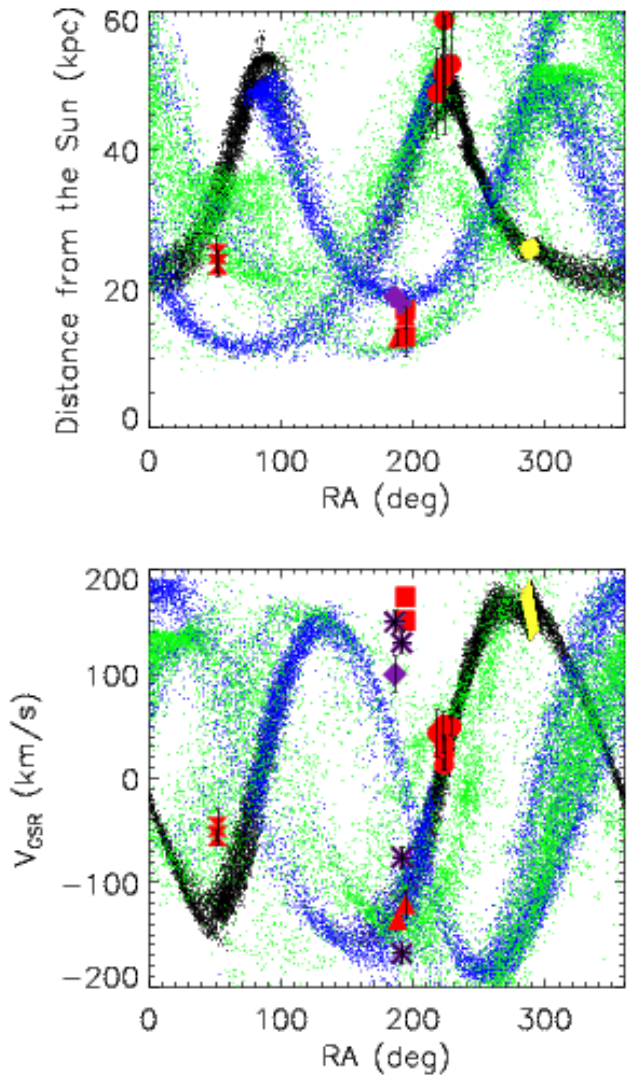}
\caption{Distance from the Sun vs. right ascension (top panel) and
radial velocity vs. right ascension (bottom panel)
for all the structures previously shown in Figures \ref{SV_polar1} and
\ref{SV_polar2} (same color coding in the online version) with known distances and
velocities. The Sagittarius model (blue, green, black, and yellow dots) is not shown as 
a fit to the Virgo substructures. \citet{new07} link a Galactic standard of rest 
velocity of 130 $\pm$ 10 km s$^{-1}$ to the S297+63-20.5 overdensity. Other velocities 
shown here with purple asterisks correspond to subsequent significant moving groups 
they found in these fields. \label{SV_distvel}}
\end{figure}

Figures \ref{SV_polar1}--\ref{SV_distvel} show
the location of the overdensities reported in the literature and of
the group and pairs we have just identified in a polar plot and in
distance and velocity respectively. Our pairs match the known Virgo
overdensities, both from SDSS and RR Lyrae surveys, relatively well in
sky position, distance and velocity. It is interesting to note that
these results suggest that Virgo is not a global `smooth' feature of
the Galaxy \citep[see also][]{xu06}. Rather it resembles a very
complex `Spaghetti bowl', because all the stars observed toward this
direction are on kinematic substructures and do not show a smooth
Gaussian-like underlying velocity distribution.

\subsection{Other matches}\label{sec_oth}

There is remarkable similarity between the properties of our pair 3
and of clump 3 from \citet{cle06}. Adding their BHB stars in that
clump to our sample and performing our 4distance measure results in a
large group in which contains both our pair 3 and three of their stars
from their clump 3.

\section{Discussion}\label{sec_discuss}

Based on the number of stars (101 giants), Spaghetti is a small
survey, certainly compared with very extensive survey projects like
SDSS. The main aspects that have made the Spaghetti project unique are
the high quality of its data, its depth--which has allowed the
discovery of giant stars out to $\sim100$ kpc, and the amount of
information for every object: distances with `just' 15\% error bars, a
thorough luminosity classification, radial velocity information
(critical for the identification of substructure) and metallicity
measures for every halo giant.

Through our newly defined distance measure, the 4distance, we have
been able to trace large substructures in the outer halo of the Milky
Way using the Spaghetti data set. We have confidently identified a
clump of debris from the disrupting Sagittarius dwarf galaxy as well
as other pairs that might be part of Sagittarius' older debris or are,
more likely, associated with Virgo substructures. These results are
found to be quite robust under small changes of the group-finding
algorithm. We tried several weighting factors as well as different
levels of the $4distance$ measure and found that both influence the
overall result only marginally.

A method which resembles ours slightly, is the Stellar Pair Technique
developed by \citet{cle06}. This requires pairs to have at most a
three-dimensional distance of 2 kpc and a radial velocity difference
of $\leq25$ km s$^{-1}$. These choices for the separation are motivated by
the suspected characteristics of streams.  They do not take into
account the varying errors in the observables, nor chance
clustering. This is why we deem our method, which by itself selects a
clustering scale, more suitable.

We have also tested the 4distance method on Monte Carlo data sets drawn
from simulations of disrupted satellites. This analysis shows that
when a substructure is identified, 76\% of its members share a common
physical origin. This further strengthens our conviction that this
method is reliable. 

While the Spaghetti survey is well suited to find and trace
substructures especially far out in the halo, like any red giant
survey it has limitations on the surface brightness of the
substructures it can detect. For example the Orphan Stream, which has
been modeled to originate from a progenitor with a stellar mass of
$\sim7.5 \times 10^5 M_{\odot}$ \citep{sal08}, seems to be right on
this boundary. While the Spaghetti survey has three fields right where
the Orphan Stream reaches its peak surface brightness, we only find two
probable candidate members in these fields \citep{sal08}.  This also
means that ``pure'' red giant surveys may not be able to put any
constraints on the low-mass end of the luminosity function of objects
accreted very early on, perhaps analogous to the recently discovered
ultrafaint satellites \citep[e.g.,][]{sim07}. On the other hand, as shown in Figures
\ref{sim_skydistrstreams} and \ref{4dist_simsurfaceB}, the only
limiting factor for detection of substructure by our 4distance method
seems to be the surface density of the stream. In this sense, the
4distance method is not biased toward accretions of any particular
type, with the caveat that the size of the data set will impose a
lower limit on the surface brightness of the structures the method
will be able to detect.

The main goal of the Spaghetti survey was to establish the amount of
substructure in the stellar halo. The analysis we have performed on
this data set allows us to quantify this combining kinematics and spatial 
information, which is a unique approach. In our 
significant pairs we found 20 giants, which is 20\% of all the giants
in the Spaghetti data set. Of these 20 giants, 19 were found in the
first step of the method, at $4dist \leq 0.05$. From the analysis of
random sets we expect about five pairs with on average nine giants
in them to be chance matches. This would leave 10 `real' matched
giants in the data set. We think this measure is conservative, because
both the amount of substructure we found which can be linked to already
known substructures, like Sagittarius and Virgo, already indicate a
fraction in substructures of $>$10\% and the analysis of the pairs
found in the simulated data sets also show that a high percentage of
all matches might be real. Finally and more importantly, even in the
simulated data sets only 25\% of the `stars' were in pairs according
to our method, despite the fact that the simulated halo was
\textit{fully} composed of streams.

Our limitation is clearly the size of the sample, which means that
Poissonian statistics dominate our analysis and estimation of the
significance of our results. The comparison to the simulations
evidences that we cannot put an upper limit on the total amount of
substructure in the halo. Although in our data set we find no more than
20\% to be in large substructures well above our detection limit,
which would be around the mass of the Orphan Stream progenitor, it is
also consistent with the whole outer halo having been built from
accreted satellites. Larger spectroscopic surveys will probably be
able to improve this significantly. We estimate using our simulations
that in samples with 1000 giants the amount of substructure detected
by the 4distance method should raise to approximately 76\%. It will be
important in the context of ongoing and future surveys (e.g., SEGUE, \citet{yan09}; SEGUE II, \citet{roc09}; HERMES, \citet{ras08}; and WFMOS, \citet{bas05}) to confirm these estimates
using cosmological simulations of stellar halos. We defer this to
future work, as well as establishing which is the best strategy to map
the halo (e.g., contiguous fields vs. pencil beams) to unravel its
assembly history.

We were able to trace the Sagittarius dwarf galaxy and the Virgo
overdensity, the only two known large substructures in the part of the
sky probed by the Spaghetti survey. However, because the survey has
only probed a small number of directions on the sky, it is very well
possible that we have missed relatively large substructures at larger
distances, or in other directions. It is remarkable however that
although $\sim$35\% of the fields observed by Spaghetti are outside of
the sky coverage of SDSS, only one pair of stars (pair 6) is found in
that region.  This would suggest that the substructure in the halo
is not isotropically distributed on the sky, and therefore is unlikely
to be the result of the overlap of a large number of high surface
brightness, narrow streams.

Most of our 30 simulated data sets show a larger amount of substructure
than found in the Spaghetti survey. On average there is an excess of
structure at small scales ($4dist \leq 0.05$). Therefore, although our
results are consistent with the whole stellar halo being built by
accretion, the characteristic size of the structures found by the
Spaghetti survey is larger than what is produced by
$10^{7}$M$_{\odot}$ satellites. This can be due to earlier accretion,
or more massive satellites. Some additional support to this
interpretation comes from the fact that the estimated mass for the
original Sagittarius galaxy is $\sim$50--100 times the mass of our
simulated satellites \citep{hel04,law05}.

\section{Conclusions}\label{sec_conc}

We have developed a method to measure the amount of substructure in
surveys consisting of spatial and radial velocity information for halo
stars. When applied to the final data set from the Spaghetti survey, we
find one group and seven pairs which contain a total of 20 stars. The most
outstanding group, with six members, can confidently be associated with
debris from the disrupting Sagittarius dwarf galaxy. Another pair
might be associated with older Sagittarius debris.  On the basis that
the Virgo overdensity is not linked to the Sagittarius leading tail
(as demonstrated by \citet{new07}), two other
pairs can be associated with known Virgo overdensities. Simulations in
which this works, prefer a prolate halo shape.

The stars in groups and pairs constitute 20\% of the Spaghetti
data set. Comparison with random sets allows us to derive a very
conservative lower limit of 10\% of the stars to be truly associated
to substructures. Unfortunately, no conclusive upper limit can be
given. From comparison with data sets drawn from a simulated halo made
entirely of $10^{7}$M$_{\odot}$ disrupted satellites we find that the
Spaghetti data set marginally supports that the whole stellar halo was
built by accretion of {\it such} galaxies. The characteristics of the
substructure found in the Milky Way halo seem to imply that broad
streams dominate our data set. This would suggest early merging and/or
relatively heavy progenitors.

Further insights and better constraints may be obtained from deep
imaging of the regions around these
substructures and from high-resolution spectroscopy of the giant stars
listed in Table 2.

\begin{acknowledgements}
We thank the referee for useful suggestions that helped improve the paper. A.H. and E.S. gratefully acknowledge the Netherlands Foundation for
Scientific Research (NWO) and the Netherlands Research School for
Astronomy (NOVA) for financial support. E.S. thanks the Astronomy 
Department of Case Western Reserve University for their warm hospitality 
and support. H.L.M. acknowledges support from NSF grants AST-0098435 and 
AST-0607518. E.O. acknowledges partial support through NSF grants AST-0098518,
0205790, 0505711, and 807498. M.M. acknowledges partial support through
NSF grants AST-0098661, 0206081, 0507453, and 0808043.\end{acknowledgements}

\appendix
\section{The Great Circle method}\label{gc_sec}

The 4distance method as developed in this work is suitable to look for substructures on small
scales; predominantly clumplike structures. On the other hand, we
also expect to see a significant amount of large-scale streamlike
structures, particularly in the outer halo. To search for these
streams, we adopt a great circle method \citep{lyn95,pal02}. The main
assumption underlying this method is that accreted debris from a
satellite orbits in a plane containing both the current position of
the satellite and the Galactic center, whose intersection with the
celestial sphere is a great circle on the sky. This assumption
requires a spherical potential, a requirement which holds relatively
well in the outer halo.

All objects on the same orbit share also the same `orbital pole',
defined by the direction of their angular momentum. For each star, the
direction of the orbital pole is perpendicular to the vector drawn
from the Galactic center to the star's current position. An indication
of \textit{possible} linkage in dynamical history in our data set would
therefore be to find several objects on a great circle associated with
(so perpendicular to) a common orbital pole. Due to the pencil-beam
character of the Spaghetti survey, however, it is not feasible to
perform an investigation based on just the sky positions (and thus great
circles) of the stars alone, like for instance a pole-count analysis
as performed in \citet{iba01c} using C stars from the APM survey.

\subsection{Specific Energy and Angular Momentum}

The specific energy of the star's motion is given by:\\
\begin{equation}
E = \case{1}{2}v_{gal}^{2}+\case{1}{2}h^{2}r^{-2}+\Psi 
\label{specE}
\end{equation}

Although the exact values of the angular momentum, $h$, and the
specific energy, $E$, of each star are unknown, we may assume that they are
constant for debris from the same parent satellite \citep{lyn95}. The
distance from the Galactic center, $r$, is available and a first
approximation to the radial velocity as seen from the Galactic Center
($v_{gal}$) is given by the measured line-of-sight velocity after
correction for the motion of the local standard of rest
($v_{GSR}$). Furthermore, we may assume a functional form for the
Galactic potential, $\Psi$. Rewriting Equation \ref{specE} as:\\
\begin{equation}
E_{r} = E-\case{1}{2}h^{2}r^{-2} \approx \case{1}{2}v_{GSR}^{2}+\Psi 
\end{equation}
we obtain a first approximation for $E_{r}$. Because $E$ and $h$ are
constants, we expect to see a linear dependence in an $E_{r}$ vs.
$r^{-2}$ plot for all the stars originating from a common parent
satellite.

\begin{figure}
\epsscale{1.00}
\includegraphics[width=\linewidth]{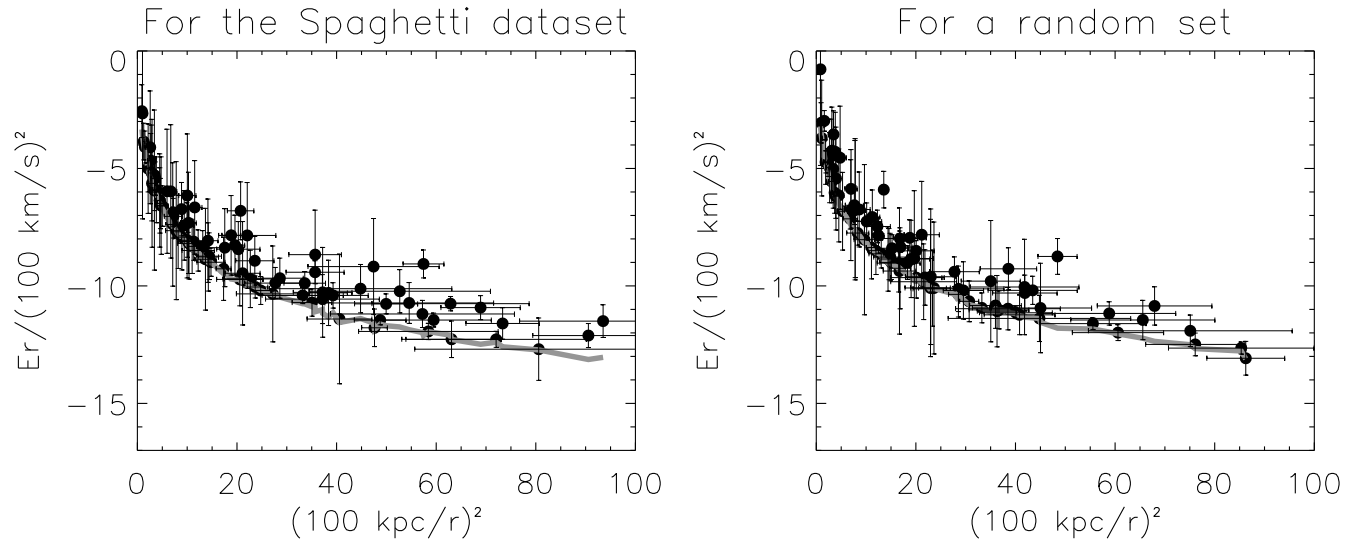}
\caption{Left panel shows the $E_{r}$ vs. $r^{-2}$ diagram for all stars 
farther out than 10 kpc in the Spaghetti data set. The error bars account for
distance and velocity uncertainties. The solid gray line represents
the contribution of the Galactic potential to $E_r$ for these
stars. The right panel shows a similar plot, but for a random set instead of 
the Spaghetti data set.\label{Erplot}}
\end{figure}

The results for all the stars that are farther out than 10 kpc in our
data set are shown in the upper panel of Figure \ref{Erplot}. The gray
solid curve indicates the contribution to $E_{r}$ of the Galactic
potential for these stars, obtained using the model of Johnston,
Spergel and Hernquist (1995). The figure shows that most of the stars
follow the trend dictated by the overall potential. The bottom panel
of Figure \ref{Erplot} shows the same diagram for a random set,
constructed as described in Section \ref{sec_binsize}. Clearly there is
no significant difference between the two panels. This leads us to
conclude that this method is not suitable for our data set. Although by
eye it appears possible to fit straight lines through many points in
the top panel, this is also the case for the bottom panel, which is
devoid of substructure by construction. Another concerns are the
extensive error bars in the data set.

\subsection{Combination of both requirements}\label{gc_com_sec}

Because of the limitations discussed above, we will only use the great
circle method in a complementary fashion to the 4distance method in
order to determine whether any of the previously found substructures
can, on the basis of sky position, energy and angular momentum
considerations, be linked to other structures located on the sky.

We measure the average position of every group or pair shown in Table
\ref{tab_final_groups} which is farther out than 10 kpc from the
Galactic center. In the search for the angular momentum pole each
group or pair is thus treated as one object. We bin the sky in 2.5$^{\circ}$
by 2.5$^{\circ}$ areas where every sky element represents an orbital
pole direction. Subsequently we count the number of objects found on a
10$^{\circ}$ wide band following the great circle of each orbital pole
direction. If three or more objects are found on one great circle, a
least-square method is used to define the likelihood that the
corresponding members of these objects in the $E_{r}$ vs. $r^{-2}$
diagram can be fitted by a straight line. We require a high
probability (Q=0.99) on the straight line fit and a small error on the
slope and intersection with the y-axis ($ \leq 10\%$). If such a fit
can be obtained, all the stars found in the groups or pairs associated with one
orbital pole direction are considered to be possible debris from the
same parent satellite. For every match the largest possible
association is considered, provided that at least three of the objects were 
initially more than 10$^{\circ}$ apart on the sky.

When applying this method to the previously found groups and pairs within the
Spaghetti data set, of which five are beyond 10 kpc, we find one
association of groups and pairs that have a possible dynamical linkage,
including group 1 and pairs 5, 7 and 8. Although in the paper, we argued that 
the stars in these group and pairs in fact might be unrelated in dynamical origin, 
belonging to separate Virgo and Sagittarius substructures (see Section 
\ref{sec_virgo}), their linkage by this method is not surprising since 
they do lie closely to a single great circle on the sky.

\subsection{Substructure in the simulated data sets: great circles}\label{sec_sim2}
We now focus on the combined great circle method, and
apply it to the five simulated data sets, described in Section
\ref{sec_sim}, which show the closest resemblance to our data set
(numbers 7, 17, 22, 23 and 25). The results for these simulated data sets are given in 
Table \ref{tab_gc_sims}. The number of associations of linked groups and pairs
with one great circle on the sky and of which their `giants' were on a
straight line in the $E_{r}$ vs. $r^{-2}$ diagram are given in the
second column. The third column shows the fractions of associations that are `correct'. 
We define an association of groups or pairs to be `correct' if at least two of
its groups or pairs are from a common progenitor and if more than 50\% of the
stars within the association originate from this common parent
satellite. 

In total only 12.5\% of all associations in these five simulated 
data sets are called `correct' using our criteria. None of the associations 
links purely stars from one common satellite. This poor result is partly due 
to the fact that almost none of the simulated data sets possess enough groups or pairs
(as defined by the 4distance method) from the same progenitor. In our
implementation of the great circle method, we can link only three or
more groups or pairs, while almost never three or more groups or pairs from the same
object are found in our simulated data sets. Only in simulated data set
17, more than two groups were found that stem from a common
progenitor. Still, the great circle method matches a substantial
number of unrelated groups and pairs together. In three of the five examined
simulated data sets, all groups and pairs were found in at least one
association. This result shows that great care must be exercised when
using the great circle method on data sets which are as small, have a
nonuniform sky distribution and as ``large'' distance errors as the
Spaghetti data set.

\begin{deluxetable}{ccc}
\tabletypesize{\footnotesize}
\tablecaption{Properties of the associations of groups and pairs found using the 
combined great circle method on the data set and several simulated data sets.
\label{tab_gc_sims}}
\tablehead{\colhead{Data Set} & \colhead{\# Associations} & \colhead{Fraction of} 
\\ & & \colhead{correct associations}}
\startdata
Spaghetti & 1 & ?/1 \\
Sim7 & 7 & 0/7 \\
Sim17 & 9 & 2/9 \\
Sim22 & 2 & 0/2 \\
Sim23 & 3 & 1/3 \\
Sim25 & 3 & 0/3 \\
\enddata
\end{deluxetable}

\subsection{Conservation of total energy and angular momentum and the
role of errors}

\begin{figure}
\epsscale{1.0}
\begin{center}
\includegraphics[width=0.7\linewidth]{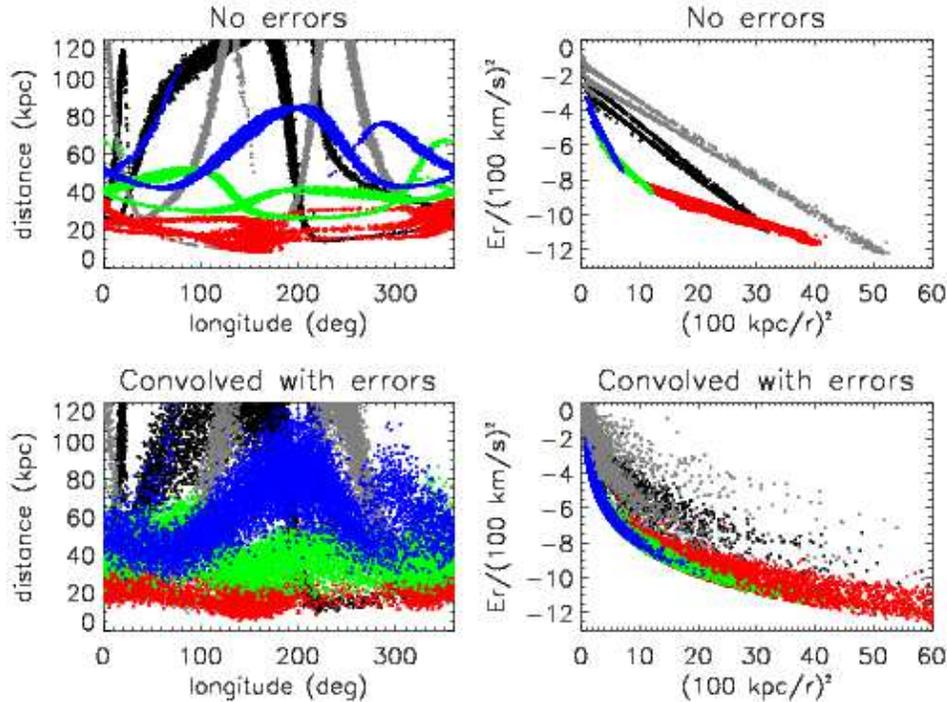}
\caption{All `giants' from simulations of five disrupted
satellites. The left panels show the streams in galactic longitude
vs. distance. In the right panels the stream `giants' are plotted
using the same color coding in an $E_{r}$ vs. $r^{-2}$ diagram. However, the
top panels use the original simulated streams and the true value
for $E_{r}$, while in the bottom panels the `giants' properties are
convolved with `observational' errors and the line-of-sight
velocities, transformed to the Galactic standard of rest, are used as
a proxy to the radial velocity of a star.\label{simclosefar}}
\end{center}
\end{figure}

We can use our simulations also to understand how well the assumption
of conservation of angular momentum holds, and what the effects of
errors (observational, but also due to projection effects) are. In
Figure \ref{simclosefar} we plot a subset of five streams from the
final output of the simulations (the same subset of streams as was 
used in Section \ref{sec_4dist_sim2}). Shown are the distance from the Sun
vs. galactic longitude (upper panel), the theoretical $E_{r}$
vs. $r^{-2}$ diagram for the simulations themselves, using no
errors and computed using the true radial velocities from the Galactic
center (middle panel) and the `observed' $E_{r}$ vs. $r^{-2}$
diagram obtained by convolving the simulations with `observational'
errors (bottom panel). As expected satellites on orbits which come
close to the Galactic center conserve their total angular
momentum (and angular momentum pole) less well as is shown in the middle
panel. However, the observational errors and our limited leverage on
the radial velocity (i.e., as measured from the Sun) are mostly
responsible for destroying the tight correlations in the $E_{r}$
vs. $r^{-2}$ diagram between particles from a common satellite.

\newpage

\begin{thebibliography}{}
\bibitem[Adelman-McCarthy et al.(2007)]{sdss07} 
Adelman-McCarthy, J.~K., et al.\ 2007, \apjs, 172, 634 
\bibitem[Bassett et al.(2005)]{bas05} Bassett, B.~A., Nichol, B., \& Eisenstein, D.~J.\ 2005, Astron. Geophys., 46, 050000 
\bibitem[Bell et al.(2008)]{bell07} Bell, E.~F., et al.\ 2008, 
\apj, 680, 295
\bibitem[Bellazzini et 
al.(2006)]{bela06} Bellazzini, M., Newberg, H.~J., Correnti, 
M., Ferraro, F.~R., \& Monaco, L.\ 2006, \aap, 457, L21 
\bibitem[Belokurov et al.(2006)]{bel06} Belokurov, V., et 
al.\ 2006, \apjl, 642, L137 
\bibitem[Belokurov et al.(2007a)]{bel07a} Belokurov, V., et 
al.\ 2007a, \apjl, 657, L89 
\bibitem[Belokurov et al.(2007b)]{bel07b} Belokurov, V., et 
al.\ 2007b, \apj, 658, 337
\bibitem[Bullock \& Johnston(2005)]{bul05} Bullock, J.~S., 
\& Johnston, K.~V.\ 2005, \apj, 635, 931 
\bibitem[Chou et al.(2007)]{cho07} Chou, M.-Y., et al.\ 2007, 
\apj, 670, 346
\bibitem[Clewley 
\& Kinman(2006)]{cle06} Clewley, L., \& Kinman, T.~D.\ 2006, \mnras, 371, L11 
\bibitem[Dehnen \& Binney(1998)]{deh98} Dehnen, W., \& Binney, J.~J.
\ 1998, \mnras, 298, 387 
\bibitem[de Jong et al.(2007)]{deJ07} de Jong, R.~S., et al.\ 
2007, IAU Symposium, 241, 503 
\bibitem[de Jong et al.(2008)]{deJ08} de Jong, R.~S., 
Radburn-Smith, D.~J., \& Sick, J.~N.\ 2008, Astronomical Society of the Pacific Conference Series, 396, 187 
\bibitem[Dohm-Palmer et al.(2000)]{doh00} Dohm-Palmer, R.~C., 
Mateo, M., Olszewski, E., Morrison, H., Harding, P., Freeman, K.~C., 
\& Norris, J.\ 2000, \aj, 120, 2496
\bibitem[Dohm-Palmer et al.(2001)]{doh01} Dohm-Palmer, R.~C., 
et al.\ 2001, \apjl, 555, L37 
\bibitem[Duffau et al.(2006)]{duf06} Duffau, S., Zinn, R., 
Vivas, A.~K., Carraro, G., M{\'e}ndez, R.~A., Winnick, R., 
\& Gallart, C.\ 2006, \apjl, 636, L97
\bibitem[Grillmair \& Dionatos(2006)]{gri06} Grillmair, C.~J., 
\& Dionatos, O.\ 2006, \apjl, 643, L17
\bibitem[Harding et al.(2001)]{har01} Harding, P., Morrison, 
H.~L., Olszewski, E.~W., Arabadjis, J., Mateo, M., Dohm-Palmer, R.~C., 
Freeman, K.~C., \& Norris, J.~E.\ 2001, \aj, 122, 1397
\bibitem[Helmi(2004)]{hel04} Helmi, A.\ 2004, \mnras, 351, 
643 
\bibitem[Helmi \& White(1999)]{hel99} Helmi, A., \& White, 
S.~D.~M.\ 1999, \mnras, 307, 495
 \bibitem[Helmi \& White(2001)]{hel01} Helmi, A., \& White, 
S.~D.~M.\ 2001, \mnras, 323, 529 
\bibitem[Ibata et al.(1994)]{iba94} Ibata, R.~A., Gilmore, 
G., \& Irwin, M.~J.\ 1994, \nat, 370, 194
\bibitem[Ibata et al.(2001a)]{iba01a} Ibata, R., Irwin, M., 
Lewis, G., Ferguson, A.~M.~N., \& Tanvir, N.\ 2001a, \nat, 412, 49
\bibitem[Ibata et al.(2003)]{iba03} Ibata, R.~A., Irwin, 
M.~J., Lewis, G.~F., Ferguson, A.~M.~N., 
\& Tanvir, N.\ 2003, \mnras, 340, L21 
 \bibitem[Ibata et al.(2001b)]{iba01b} Ibata, R., Irwin, M., 
Lewis, G.~F., \& Stolte, A.\ 2001b, \apjl, 547, L133
\bibitem[Ibata et al.(2001c)]{iba01c} Ibata, R., Lewis, G.~F., 
Irwin, M., Totten, E., \& Quinn, T.\ 2001c, \apj, 551, 294 
\bibitem[Ibata et al.(2007)]{iba07} Ibata, R., Martin, N.~F., 
Irwin, M., Chapman, S., Ferguson, A.~M.~N., Lewis, G.~F., 
\& McConnachie, A.~W.\ 2007, \apj, 671, 1591
\bibitem[Ivezi{\'c} et al.(2000)]{ive00} Ivezi{\'c}, {\v Z}., 
et al.\ 2000, \aj, 120, 963 
\bibitem[Johnston et al.(2008)]{joh08} Johnston, K.~V., 
Bullock, J.~S., Sharma, S., Font, A., Robertson, B.~E., 
\& Leitner, S.~N.\ 2008, \apj, 689, 936 
\bibitem[Johnston et al.(1996)]{joh96} Johnston, K.~V., 
Hernquist, L., \& Bolte, M.\ 1996, \apj, 465, 278 
\bibitem[Johnston et al.(2001)]{joh01} Johnston, K.~V., 
Sackett, P.~D., \& Bullock, J.~S.\ 2001, \apj, 557, 137 
\bibitem[Johnston et al.(1995)]{joh95} Johnston, K.~V., 
Spergel, D.~N., \& Hernquist, L.\ 1995, \apj, 451, 598
\bibitem[Juri{\'c} et al.(2008)]{jur08} Juri{\'c}, M., et 
al.\ 2008, \apj, 673, 864
\bibitem[Keller et al.(2008)]{kel08} Keller, S.~C., Murphy, 
S., Prior, S., DaCosta, G., \& Schmidt, B.\ 2008, \apj, 678, 851
\bibitem[Law et al.(2005)]{law05} Law, D.~R., Johnston, 
K.~V., \& Majewski, S.~R.\ 2005, \apj, 619, 807
\bibitem[Lynden-Bell \& Lynden-Bell(1995)]{lyn95} Lynden-Bell, D., 
\& Lynden-Bell, R.~M.\ 1995, \mnras, 275, 429
\bibitem[Majewski et al.(2003)]{maj03} Majewski, S.~R., Skrutskie, 
M.~F., Weinberg, M.~D., 
\& Ostheimer, J.~C.\ 2003, \apj, 599, 1082 
\bibitem[Martin et al.(2004)]{martin04} Martin, N.~F., Ibata, 
R.~A., Bellazzini, M., Irwin, M.~J., Lewis, G.~F., 
\& Dehnen, W.\ 2004, \mnras, 348, 12 
\bibitem[Mart{\'{\i}}nez-Delgado et al.(2001)]{mar01} 
Mart{\'{\i}}nez-Delgado, D., Aparicio, A., G{\'o}mez-Flechoso, M.~{\'A}., 
\& Carrera, R.\ 2001, \apjl, 549, L199 
\bibitem[Mart{\'{\i}}nez-Delgado et al.(2008)]{mar08} 
Mart{\'{\i}}nez-Delgado, D., Pe{\~n}arrubia, J., Gabany, R.~J., Trujillo, 
I., Majewski, S.~R., \& Pohlen, M.\ 2008, \apj, 689, 184 
\bibitem[Mart{\'{\i}}nez-Delgado et al.(2009)]{mar09} 
Mart{\'{\i}}nez-Delgado, D., Pohlen, M., Gabany, R.~J., Majewski, S.~R.,
 Pe{\~n}arrubia, J., \& Palma, C.\ 2009, \apj, 692, 955
\bibitem[Mathewson et al.(1974)]{mat74} Mathewson, D.~S., 
Cleary, M.~N., \& Murray, J.~D.\ 1974, \apj, 190, 291  
\bibitem[Monaco et al.(2007)]{mon07} Monaco, L., Bellazzini, 
M., Bonifacio, P., Buzzoni, A., Ferraro, F.~R., Marconi, G., 
Sbordone, L., \& Zaggia, S.\ 2007, \aap, 464, 201 
\bibitem[Moody et al.(2003)]{moo03} Moody, R., Schmidt, B., 
Alcock, C., Goldader, J., Axelrod, T., Cook, K.~H., \& Marshall, S.
\ 2003, Earth Moon and Planets, 92, 125
\bibitem[Morrison et al.(2000)]{mor00} Morrison, H.~L., 
Mateo, M., Olszewski, E.~W., Harding, P., Dohm-Palmer, R.~C., Freeman, 
K.~C., Norris, J.~E., \& Morita, M.\ 2000, \aj, 119, 2254 
\bibitem[Morrison et al.(2001)]{mor01} Morrison, H.~L., 
Olszewski, E.~W., Mateo, M., Norris, J.~E., Harding, P., Dohm-Palmer, 
R.~C., \& Freeman, K.~C.\ 2001, \aj, 121, 283 
\bibitem[Morrison et al.(2003)]{mor03} Morrison, H.~L., et 
al.\ 2003, \aj, 125, 2502 
\bibitem[Newberg et al.(2007)]{new07} Newberg, H.~J., Yanny, 
B., Cole, N., Beers, T.~C., Re Fiorentin, P., Schneider, D.~P., 
\& Wilhelm, R.\ 2007, \apj, 668, 221 
\bibitem[Newberg et al.(2002)]{new02} Newberg, H.~J., et al.\ 
2002, \apj, 569, 245
\bibitem[Palma et al.(2002)]{pal02} Palma, C., Majewski, 
S.~R., \& Johnston, K.~V.\ 2002, \apj, 564, 736
\bibitem[Pe{\~n}arrubia et al.(2005)]{pen05} Pe{\~n}arrubia, 
J., et al.\ 2005, \apj, 626, 128 
\bibitem[Raskin \& Van Winckel(2008)]{ras08} Raskin, G., \& Van Winckel, H.\ 2008, \procspie, 7014, 70145D  
\bibitem[Rockosi et al.(2009)]{roc09} Rockosi, C., Beers, 
T.~C., Majewski, S., Schiavon, R., Eisenstein, D., 
\& with input from the SDSS-III Collaboration 2009, arXiv:0902.3484 
\bibitem[Sales et al.(2008)]{sal08} Sales, L.~V., et al.
\ 2008, \mnras, 389, 1391
\bibitem[Shang et al.(1998)]{sha98} Shang, Z., et al.\ 1998, 
\apjl, 504, L23 
\bibitem[Simon \& Geha(2007)]{sim07} Simon, J.~D., \& Geha, M.\ 2007, \apj, 670, 313 
\bibitem[Sirko et al.(2004)]{sir04} Sirko, E., et al.\ 2004, 
\aj, 127, 899 
\bibitem[Skrutskie et al.(2006)]{2mass06} Skrutskie, M.~F., et 
al.\ 2006, \aj, 131, 1163
\bibitem[Tolstoy et al.(2004)]{tol04} Tolstoy, E., et al.\ 
2004, \apjl, 617, L119 
\bibitem[Vivas et al.(2001)]{viv01} Vivas, A.~K., et al.\ 
2001, \apjl, 554, L33 
\bibitem[Vivas et al.(2004)]{viv04} Vivas, A.~K., et al.\ 
2004, \aj, 127, 1158 
\bibitem[Xu et al.(2006)]{xu06} Xu, Y., Deng, L.~C., 
\& Hu, J.~Y.\ 2006, \mnras, 368, 1811
\bibitem[Yanny et al.(2009)]{yan09} Yanny, B., et al.
\ 2009, \aj, 137, 4377
 \bibitem[Yanny et al.(2000)]{yan00} Yanny, B., et al.\ 2000, 
\apj, 540, 825 
\bibitem[Zinn et al.(2004)]{zin04} Zinn, R., Vivas, A.~K., 
Gallart, C., \& Winnick, R.\ 2004, Satell. Tidal Streams, 327, 92 
\end{thebibliography}
\end{document}